\documentclass{emulateapj}
\usepackage{natbib}
\usepackage{textcomp}
\usepackage{amsmath}
\usepackage{graphicx}

\shorttitle{GRB 121027A}
\shortauthors{Peng et al.} \slugcomment{}
\begin{document}
\title{GRB 121027A: long-lasting, energetic X-ray flares and clues to radiation mechanism
and progenitor star}
\author{Fang-kun Peng\altaffilmark{1}, You-Dong Hu\altaffilmark{1}, Shao-Qiang Xi\altaffilmark{1}, Xiang-Gao Wang\altaffilmark{1}, Rui-Jing Lu\altaffilmark{1}, En-Wei Liang\altaffilmark{1,2} and Bing Zhang\altaffilmark{3,1}}
\altaffiltext{1}{Department of Physics and GXU-NAOC Center for Astrophysics and Space Sciences, Guangxi University, Nanning, Guangxi 530004, China; lew@gxu.edu.cn}
\altaffiltext{2}{National Astronomical Observatories, Chinese Academy of Sciences, Beijing 100012, China}
\altaffiltext{3}{Department of Physics and Astronomy, University of Nevada Las Vegas, Las Vegas, NV 89154, USA}
\begin{abstract}
GRB 121027A is un-usual with its extremely long-lasting, energetic X-ray flares. The total energy
release in X-ray flares is about one order of magnitude higher than prompt $\gamma$-rays, making
it special from most long GRBs.
We show that while the prompt gamma-ray emission satisfies the empirical
$E_{\rm iso}-E_{\rm p}$ relation of typical long GRBs, the X-ray flares, whose spectra
can be fit with a cutoff-power-law model with well-constrained $E_p$, significantly
deviate from such a relation.
Nonetheless, a time-resolved spectral analysis of X-ray flares suggest that the X-ray
emission is consistent with the $L_{\rm iso}-E_{\rm p}$ relation of long GRBs.
We constrain the minimum Lorentz factor of the X-ray flares to be
$\sim 14$, which is consistent with the $\Gamma-L_{\rm iso}$ relation.
Our results imply that prompt $\gamma$-ray emission and late X-ray
flares share the similar radiation mechanism, but originate from the outflows
with different Lorentz factors.
We search for similar GRBs from the {\em Swift} GRB archives,
and find that the $z=6.29$ GRB 050904
is a carbon copy of GRB 121027A.
The long-lasting, energetic X-ray flares in these GRBs demand significant accretion at late
times, which point towards a large-radius progenitor star.
\end{abstract}
\section{Introduction}
\label{sec:intro}
Observations with {\em Swift} misison has greatly improved our understanding on the nature of the gamma-ray burst (GRB) phenomenon. The rapid slewing capacity of the X-ray telescope (XRT) onboard {\em Swift} makes it possible to catch X-ray emission from very early to late epochs of GRBs. A large sample of GRB X-ray lightcurves have been collected in the time window from tens of seconds to days or even months post the Burst Alert Telescope (BAT) triggers. A canonical X-ray lightcurve with five components was revealed from the sample (Zhang et al. 2006; Nousek et
al. 2006)\footnote{The X-ray lightcurves of a small fraction of GRBs
are a featureless single power-law (Liang et al. 2009; Evans et al. 2009).}.

X-ray flares are observed for about half of the $\rm Swift$
GRBs (Burrows et al. 2005; Falcone et al. 2007; Chincarini et al. 2007). Most flares
happened at $t<1000$ seconds and a small fraction of flares occurred
at $\sim 10^5$ seconds post the GRB trigger time ($T_{0}$). No
significant flare after the prompt gamma-ray phase was detected for
about 1/3 of these GRBs, but bright X-ray flares were detected for the
other 2/3 of GRBs (Qin et al.
2013). Joint spectral analyses for the prompt gamma-ray emission and early
X-ray flares indicate that X-ray flares are the low-energy extension
of the prompt gamma-ray emission (Peng et al. 2013).
Significant flares that dominate late X-ray emission
(without the detection of the afterglow emission up to $\sim 10^5$ seconds)
were observed in some GRBs, such as GRBs 050502B, 050724, 050904,
and 060223. The most prominent one is GRB 050904 (Cusumano et al. 2007),
which was suggested as a super-long GRB (Zou et al. 2006).
It is generally believed that
these X-ray flares are due to extended central engine activity at
late times (Burrows et al. 2005; Zhang et al. 2006; Fan \& Wei 2005;
King et al. 2005; Dai et al. 2006; Perna et al. 2006; Proga \&
Zhang 2006; Liang et al. 2006; Lazzati \& Perna 2007).
Late flares may be critical for revealing the global evolution of the GRB
central engine, hence the properties of the progenitor stars.
It is unclear whether the X-ray flares share the relations between
spectral peak energy ($E_{\rm p}$) and burst energetics that have been found for
long GRBs with the prompt $\gamma$-ray data, such as the
$E_{\gamma, \rm iso}-E_{\rm p}$ relation (the Amati-relation;
Amati et al. 2002) and the $L_{\gamma, \rm iso}-E_{\rm p}$ relation
(Yonetoku et al. 2004; Liang et al. 2004), where $E_{\gamma, \rm iso}$
(or $L_{\rm \gamma, iso}$) is the bolometric isotropic gamma-ray energy
release (or luminosity) in a wide band ($1-10^4$ keV).

{\em Swift}/BAT triggered an unusual GRB on Oct. 27, 2012, which has
extremely bright and long flares that last up to $\sim 10^{5}$ s
post the GRB trigger (Evans et al. 2012).  Here we present an analysis of
GRB 121027A in order to study whether the late flares share the same
relations with the prompt gamma-rays.
The concordance cosmology with parameters $H_0 = 71$ km s$^{-1}$
Mpc$^{-1}$, $\Omega_M=0.30$, and $\Omega_{\Lambda}=0.70$ is
adopted to calculate burst energetics.

\section{Data Analysis Results}\label{sec:using}
GRB 121027A
has a redshift 1.773 (Levan et al. 2012). We downloaded the BAT and XRT data
from the {\em Swift}
archive\footnote{http://heasarc.nasa.gov/docs/swift/archive/}, and
extracted the lightcurves and spectra in the BAT and XRT bands. The
0.3-10 keV band lightcurve is shown in Fig. 1, with XRT data directly
plotted and the BAT data extracted to this energy band based on its
spectral information in the BAT band.
The duration $T_{90}$ in the BAT band is $62.6\pm 4.8$
seconds. A steep decay segment from $T_0+70$ to $T_0+180$ seconds is
observed in the XRT band. Bright X-ray flares are observed during
$T_0+180$ and $T_0+4\times 10^4$ seconds. The first flare peaks at around $T_0+270$ seconds. After a dip, the flux rises rapidly with a slope $\alpha>10$. The peak of the 2nd flare was not detected because of the orbit constraint. The MAXI/GSC nova alert system triggered on a
bright un-catalogued X-ray transient source with 4-10 keV flux about 150 mCrab at $T_0+2400$ seocnds, and the transient was identified as the X-ray emission from GRB 121027A (Serino et al. 2012). We correct the flux to the 0.3-10 keV energy band and also show it in Fig. 1. The third flare peaks around $T_0+7\times 10^3$ seconds. The flux from $T_0+5300$ to $T_0+5600$ seconds almost keeps constant, which maybe due to superposition of the tail of the second flare and the rising segment of the third flare. We make an empirical fit to the three flares with three broken power-law components, as shown in Fig. 1. One can observe that the peak luminosities of the three flares do not show a clear evolution pattern. After the flaring phase, a plateau between $T_0+4\times 10^4$ and $T_0+1.5\times 10^5$ seconds is observed. It
transits to a decay segment with a slope $\alpha=-1.55$ up to $3\times 10^{6}$ seconds.

The BAT spectrum is well fit with a single power-law, which yields a
photon spectral index $\Gamma_{\rm BAT}=1.97\pm 0.07$ and a reduced
$\chi^2=40/56$ dof. XRT detected the corresponding X-rays of prompt emission
from $T_0+57$ to $T_0+66$ seconds in the slewing mode. We make a joint
spectral analysis of the spectrum accumulated with BAT and XRT in
the prompt emission phase, and find that the spectrum is well fit
with a cutoff power-law, with low energy photon index $\Gamma_{\gamma}=1.49^{+0.11}_{-0.12}$
and cutoff energy $E_{\rm c}= 93^{+56}_{-27}$ keV (corresponding to an $E_{\rm p}=47^{+30}_{-17}$ keV), as shown in Fig. 2(a).

We perform a time-resolved spectral analysis of the XRT data. In our spectral fitting, the absorption of our Galaxy is fixed as $N_{\rm H}=0.015\times10^{22}$ cm$^{-2}$. In order to avoid artificial $N_{\rm H}$
variations for the GRB host galaxy caused by the intrinsic spectral evolution that commonly
observed in the brightest X-ray flares (Butler \& Kocevski 2007),
we fix the host galaxy absorption to $N^{\rm host}_{\rm
H}=1.26\times10^{22}$ cm$^{-2}$, which is derived from fitting the
late time spectrum between $T_0+16347$ and $T_0+76223$ s. We find that although a single
power-law model can present acceptable fits for the time-resolved spectra during the
flares, a cut-off power-law model significantly improves the spectral fits for some time intervals of the flares. We compare our spectral fit results with the two models in Table 1. One example of our spectral fit with an absorbed cutoff power-law model is shown in Fig. 2(b). We adopt the cut-off power-law model fits, if their C-statistic is smaller than that of the single power-law fits with $\sim 20$, as marked with a $``\surd"$ in Table 1.

The $E_{\rm p}$ or power-law photon index ($\Gamma$) evolution is shown in Fig. 1. One can see that $E_{\rm p}$ rapidly decays from
$47\pm 13$ keV down to $\sim 1$ keV during the steep decay phase of the last pulse,
being well consistent with the tail emission of the prompt prompt emission
as observed in other GRBs (e.g. Zhang et al. 2007). The time-resolved spectra of X-ray
afterglows after $t>4\times 10^5$ are well fit with an absorbed
power-law with photon spectral index $\sim (2.2-2.5)$,
being well consistent with most other GRBs (e.g. Liang et al. 2007).
\begin{figure}
\centering
\includegraphics[angle=0,scale=0.4]{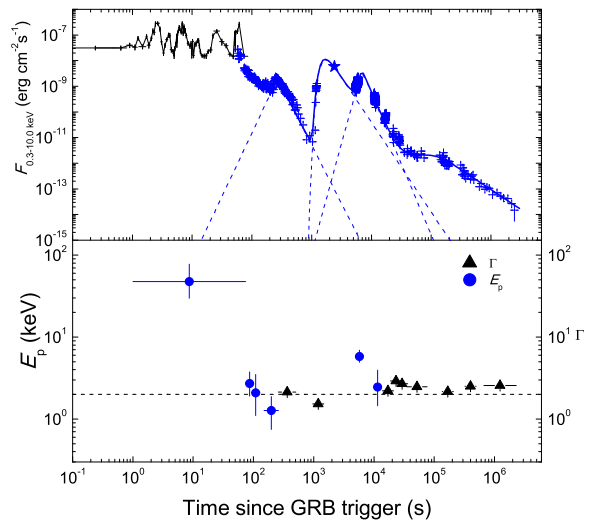}
\caption{Lightcurve in the 0.3-10 keV band and $E_{\rm p}$  or power-law photon index ($\Gamma$) evolution of GRB 121027A  derived from observations with BAT and XRT. The empirical fit for lightcurve of the late flares are also shown. Dashed lines are fits to individual flares and the solid line represents the sum of all flares. The horizonal dashed line marks $\Gamma=2$.}
\end{figure}
\section{Empirical relations}
We correct the observed $E_{\rm \gamma,iso}$ to the $1-10^4$ keV band with the spectral parameters
reported above. We obtain $E_{\gamma, \rm iso}=(2.86\pm 0.23)\times
10^{52}$ erg, where the error is calculated by a boostrap method
assuming the error of photon index $\Gamma$ has a normal
distribution and the errors of both $E_{\rm p}$ and $S_{\gamma}$ have
log-normal distributions.

In Fig. 3(a), we plot GRB 121027A in the $E_{\rm \gamma, iso}-E_{\rm p}$ diagram and
find that it well follows the empirical Amati-relation.
We then test if the flares at $t>T_0+200$ seconds are also consistent with
this relation. We calculate the X-ray fluence in the 0.3-10 keV by integrating
the X-ray flux over three flares with the spectral parameters derived above.
We obtain an X-ray fluence $S_{\rm X}=2.51\times 10^{-5}$ erg cm$^{-2}$,
which corresponds to an isotropic X-ray energy
$E_{\rm X, iso}=1.89\times 10^{53}$ erg. Note that we do
not correct the X-ray energy to the $1-10^4$ keV band, since the 0.3-10 keV
band spectrum can be well fit with a cutoff power-law, so that fluence
above $E_p$ is negligible. We can see that the total energy of the flares
is about one order of magnitude larger than the
gamma-ray energy, but $E_{\rm p}$ values of the X-ray flares are much lower
than those of gamma-rays. As shown in Fig. 3(a), the flares significantly deviate
from the Amati-relation. Next, with a time-resolved spectral
analysis, we plot $L_{\rm X, iso}$ as a function of $E_{\rm p}$ in Fig. 3(b).
It is interesting to see that the time-resolved $L_{\rm X, iso}$ and $E_{\rm p}$
are consistent with the $L_{\rm iso}-E_p$ correlation for long GRBs
(Yonetoku et al. 2004; Liang et al. 2004). These results suggest that
the radiation mechanism of prompt gamma-rays and late X-rays
is likely the same, even though the luminosity and energetics are noticeably
different.

After the flare phase, a long-lasting X-ray plateau was observed during $T_{0}+4\times 10^{4}-T_0+1.3\times 10^{5}$. This is likely the emission from the external forward shock, which is constantly refreshed by the new ejected materials that power the X-ray flares.
The end of the plateau marks the end of this energy injection, which is defined by the lower limit of the Lorentz factor of the late ejecta. This can be estimated by (Zhang et al. 2006)
\begin{equation}
\Gamma_{\rm m}=23E_{\rm X, iso,52}^{1/8}n^{-1/8}t_{\rm b,4}^{-3/8}[(1+z)/2]^{3/8},
\end{equation}
which gives $\Gamma_{\rm m}\sim 14$ for GRB 121027A.
This minimum Lorentz factor, even though violates the $\Gamma-E_{\rm iso}$ correlation (Liang et al. 2010), is consistent with the $\Gamma-L_{\rm iso}$ correlation (L\"u et al. 2012) within $2\sigma$ confidence level. Lei et al. (2013) have shown that this correlation is more fundamental, and is directly connected to the GRB central engine physics.

\begin{figure}
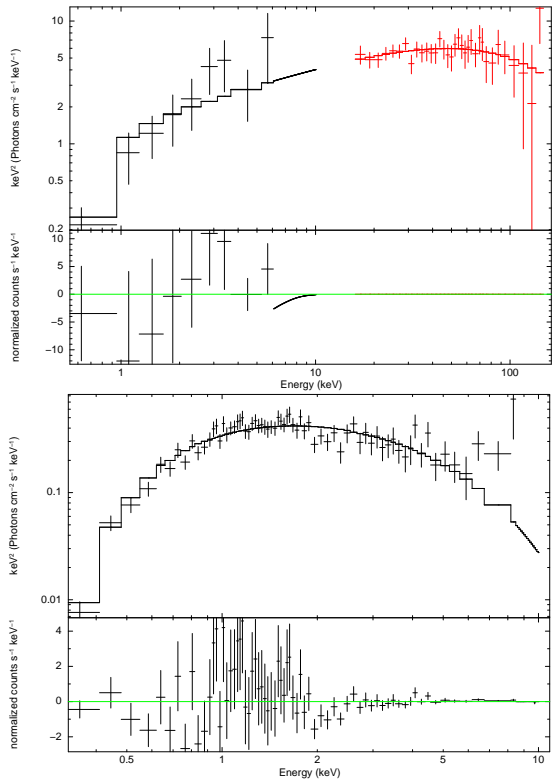

\centering
\includegraphics[angle=-90,scale=0.3]{fig2a.eps}
\includegraphics[angle=-90,scale=0.3]{fig2b.eps}
\caption{ \emph{top panel}: Joint spectral fit for the data observed with BAT and XRT for the prompt gamma-rays. \emph{bottom panel}: Example of our absorbed cut-off power-law model fit to the spectrum of the flare data in time interval $T_{0}+150\sim 260$ s.}
\end{figure}

\begin{figure}
\centering
\includegraphics[angle=0,scale=0.3]{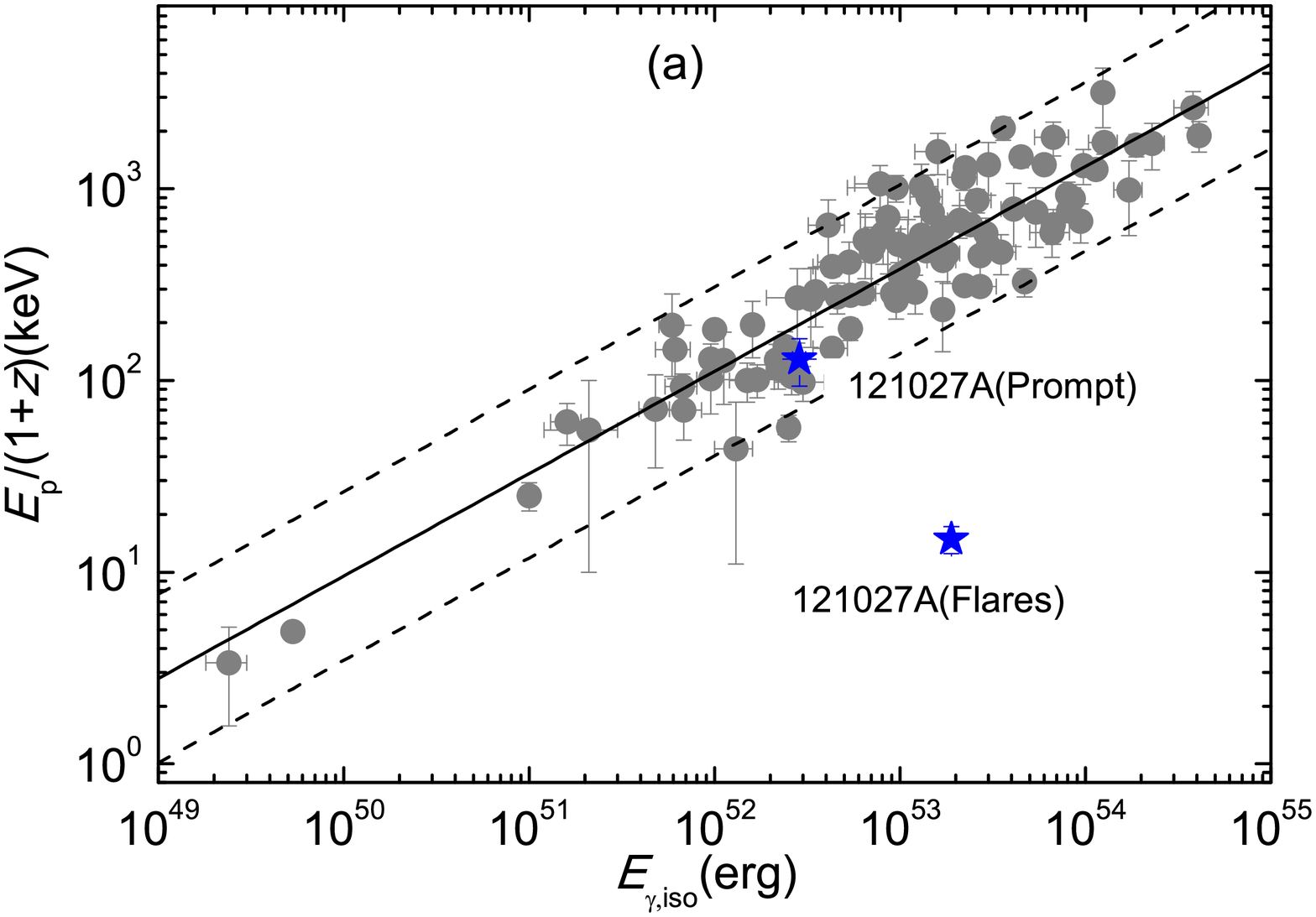}
\includegraphics[angle=0,scale=0.32]{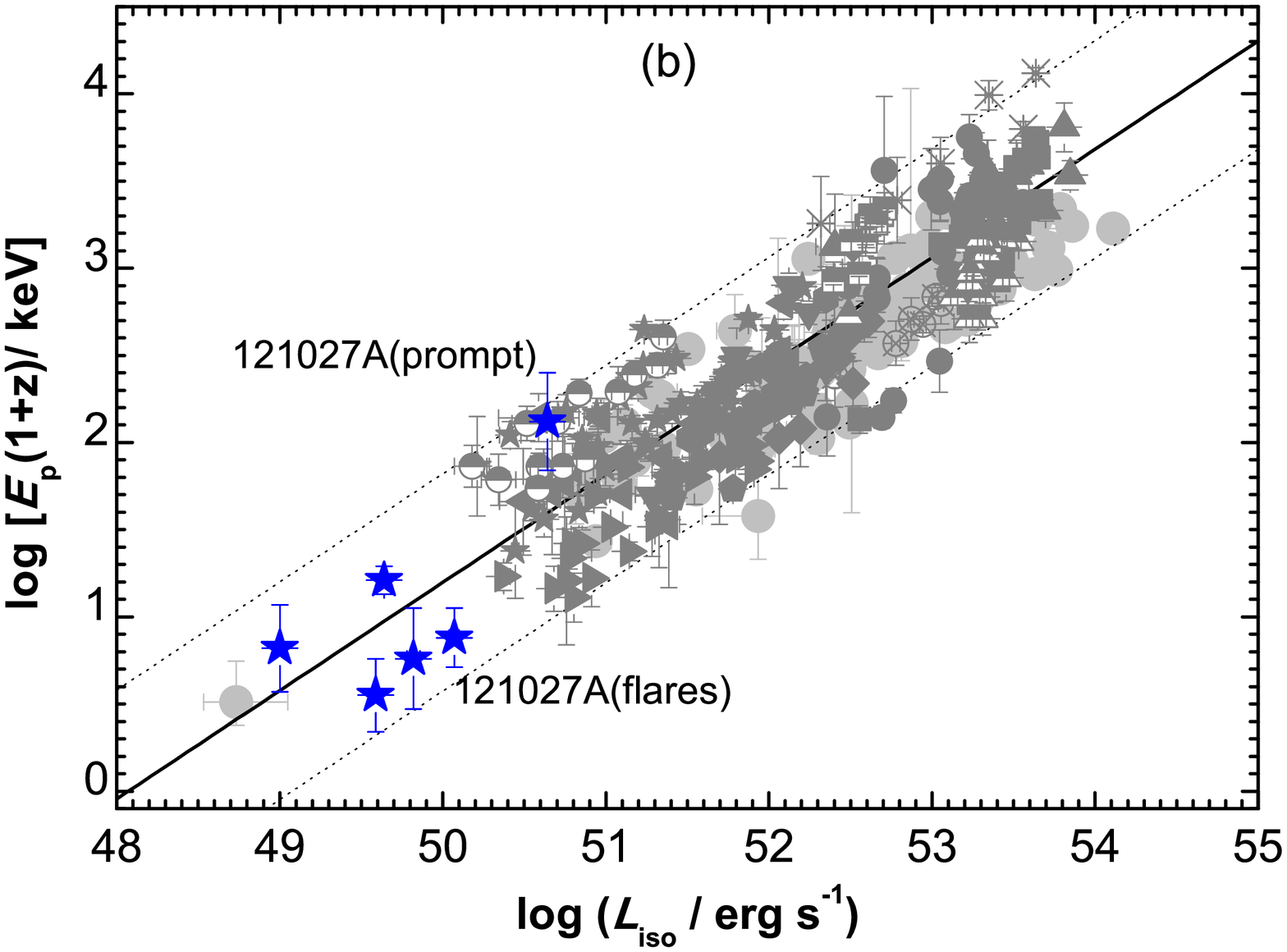}
\caption{GRB 121027A prompt gamma-ray and X-ray flares against empirical correlations of other long GRBs: (a) the Amati-relation; (b) the time-resolved $L_{\rm \gamma, iso}-E_{\rm p}$ relation. The data of the typical GRBs (grey dots) are from Amati et al. (2009), Lu et al. (2012) and references therein. Lines are the best fits and the dashed lines mark the 2 $\sigma$ region of the correlations.}
\end{figure}

\section{Physical Implications on radiation mechanism and progenitor}
\label{sec:frontmatter}

The fact that X-ray flares follow the same $L_{\rm iso}-E_p$ and $\Gamma-L_{\rm iso}$
relations as prompt emission suggest that the radiation mechanism of X-ray flares
and prompt gamma-ray emission is similar. The difference is that the more powerful
prompt emission has a higher Lorentz factor than X-ray flares.
The radiation mechanism of prompt emission
is not clear. In the literature, competing models include synchrotron radiation
(M\'esz\'aros et al. 1994; Lloyd \& Petrosian 2000; Zhang \& Yan 2011)
and quasi-thermal emission from the fireball photosphere (e.g. Thompson 1994;
Rees \& M\'esz\'aros 2005; Beloborodov 2010; Lazzati \& Begelman 2010).
At a low accretion rate that is relevant to X-ray flares, the
neutrino-anti-neutrino annihilation mechanism would not be adequate to power
X-ray flares, and magnetic dissipation is favored (e.g. Fan et al. 2005).
As a result, synchrotron emission from a magnetized jet may be the common
mechanism to power both prompt emission and X-ray flares.

Long GRBs are widely believed to be associated with the
deaths of massive stars (Colgate 1974; Woosley 1993).
The associations of some GRBs with Type Ic supernovae (Woosely \& Bloom 2006)
favor a small-size progenitor star, likely a Wolf-Rayet (WR) star with striped
hydrogen and helium envelopes. A successful jet is possible only when the
accretion time scale is longer than the time scale for the jet to penetrate
through the stellar envelope (MacFadyen \& Woosley 1999; Waxman \& M\'esz\'aros
2003; Bromberg et al. 2012). As a result, collapsar-related GRBs are typically
``long''. On the other hand, a long duration does not necessarily refer to
a large progenitor, since various mechanisms (e.g. fragmentation or magnetic
barrier, King et al. 2005; Perna et al. 2006; Proga \& Zhang 2006; Liu et al. 2012) can halt
accretion temporarily and extend the total duration of accretion. Indeed,
when X-ray flares are included, the central engine activity time of many GRBs
can be much longer than the duration of gamma-ray emission itself (e.g.
Burrows et al. 2005; Qin et al. 2013). GRB 120727A
presents an extreme case of such a long-lasting central engine activity:
Not only the X-ray flaring activity extends to a much later time, but the
total energetics of the flares is one order of magnitude higher than that
of prompt emission. This demands that a significant amount of mass is
accreted to the black hole in a later epoch. To reserve a large mass
reservoir at later times, one would demand a large progenitor star
than WRs. Since an un-mixed star would have an onion structure and
would not store a large mass in the outer layer, this burst would also
favor a progenitor star with a mixed envelope (Woosley \& Heger 2006),
as expected for a low-metallicity, rapidly-rotating star. Accretion of such
a mixed envelope tends to be smooth (Kumar et al. 2008; Perna \& MacFadyen
2010), but fragmentation or magnetic barrier (Perna et al.
2006; Proga \& Zhang 2006; Liu et al. 2012) can modulate the accretion rate to power
flares. Indeed a detailed fall-back model (Wu et al. 2013) can interpret
the X-ray lightcurve of this burst well.

We search for similar GRBs from the {\em Swift} GRB archive.
We find that the late XRT lightcurves of some GRBs are
dominated by flares without the detection of power-law afterglow component
up to $\sim 10^5$. These include GRBs 050502B, 050724, 050904, 050916, and 060223.
The most prominent one is GRB 050904 (Cusumano et al. 2007). We compare the
lightcurve of GRB 050904 with GRB 121027A in Fig. 4. One can observe that
it is almost a carbon copy of GRB 121027A.
These bursts may also have a large-size progenitor as GRB 121027A. A large
progenitor was also suggested by Levan et al. (2013), who connect GRB 121027A
with several other ultra-long GRBs, such as GRB 101225A and GRB 111209A.
We note that many more GRBs have comparable or even longer duration of
central engine activity (e.g. Qin et al. 2013). It is the energetics of the
flares that matters the most in favor of a large progenitor.

The X-ray flux up to $T_0+3\times 10^6$ seconds ($\sim 34$ days post the GRB trigger) decays as $t^{-1.55}$ after the plateau. The spectral index of the X-rays is $\beta\sim 1.5$.  This is roughly consistent with the expectation of the standard external shock model in the slow cooling regime for the wind medium case. No jet break is observed in the X-ray lightcurve. We place a lower limit of the half-opening angle ($\theta_{\rm j}$) of the jet.
In the wind medium, one has $\theta_{\rm j} > 0.202[t_{\rm j,
d}/(1+z)]^{1/4}\times (A\eta/E_{\rm X,iso,52})^{1/4}$,
where $t_{\rm j,d}=t_{\rm j}/1$ day, $\eta=0.2$ (the radiative efficiency), $E_{\rm
X,iso,52}=E_{\rm X,iso}/10^{52}$, and $A$ is wind medium parameter for a profile $n=5\times 10^{11} Ar^{-2}$ g cm$^{-3}$ (e.g., Firmani et al. 2006). We take $A=1$ that corresponds to a wind mass loss rate $\dot{M}_{\rm w}=10^{-5}M_\odot {\rm yr}^{-1}$ and a wind velocity $\dot{v}_w=10^3\ {\rm km\ s}^{1}$. The derived $\theta_{\rm j}$ is $>10.4^{\rm o}$, which is roughly consistent with that derived from the jet breaks observed in the optical bands (e.g., Bloom et al. 2003).
\begin{figure}
\centering
\includegraphics[angle=0,scale=0.3]{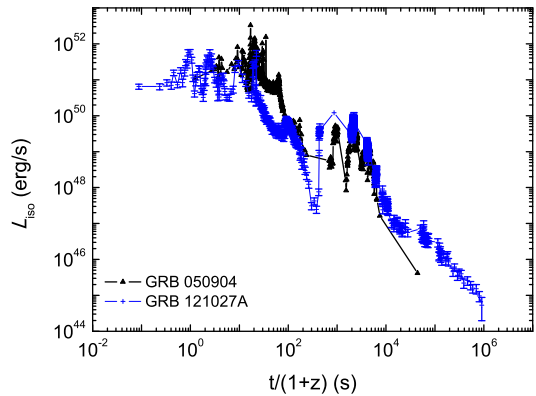}
\caption{Comparison of the lightcurves of GRB 121027A with GRB 050904 in the 0.3-10 keV band in the rest frame. }
\end{figure}

\section{Conclusion}
We have presented a temporal and spectral analysis of GRB 121027A
observed with BAT and XRT. While the prompt emission duration as detected by
BAT is only $62.6\pm4.8$ s, extremely bright flares in the XRT band last up
to $\sim 10^{5}$ s post the GRB trigger. Beyond, the
external shock was observed up to more than 34 days without detection of
a jet break.

A special property of GRB 121027A is that the total energy of late X-ray flares is one
order of magnitude higher than that of prompt emission. Yet, the radiation properties of X-ray
flares seem to follow the same $L_{\rm iso}-E_p$ and $\Gamma - L_{\rm iso}$ empirical
correlations with prompt gamma-ray emission, suggesting the same radiation mechanism
(likely synchrotron) in the two phases. The large energy budget at a late epoch demands
a large-size progenitor star, probably with mixed envelope layers. These progenitors may
be different from the WR progenitor invoked to interpret most long GRBs.
These GRBs form a distinct sub-category of long GRBs, another prototype of which is
GRB 050904. These results suggest that the GRB progenitors can be diverse, not limited
to a narrow channel. Future observations of possible associated SNe of these sub-category
of GRBs would be of great value to infer the exact nature of their progenitor stars.

\section*{Acknowledgments}
We thank helpful discussion with Xue-Feng Wu and Zi-Gao Dai. This work
made use of data supplied by the UK Swift Science Data
Centre at the University of Leicester. This work is supported by
the ``973" Program of China (2009CB824800), the National Natural
Science Foundation of China (Grants No. 11025313 and 11063001), Special Foundation for Distinguished
Expert Program of Guangxi, the Guangxi Natural Science Foundation
(2010GXNSFA013112, 2010GXNSFC013011, and Contract
No. 2011-135).
BZ acknowledges support from NSF (AST-0908362).

\begin{table}
\scriptsize
\centering
\caption{Comparison of spectral fit results with the absorbed cutoff power law model and the single power-law model for GRB 121027A. The results of models with ``$\surd$'' are accepted for our analysis.}
\begin{tabular}{llccclcc}
\hline
Interval(s) &  model & $\Gamma_{\rm c}$ & $E_{\rm c} $(keV) & C-Stat/dof & model & $\Gamma$ & C-Stat/dof  \\
\hline
1-74   &   	cpl($\surd$)	&$1.49^{+0.11}_{-0.12}$&$93.30^{+55.82}_{-27.14}$&88.7/490& pl& $1.74^{+0.05}_{-0.06}$&111.2/491\\
77-97  &	cpl($\surd$)	&$0.66^{+0.27}_{-0.28}$	&$2.03^{+0.67}_{-0.44}$	&350.3/489& pl & 1.71$\pm$0.08 &  404.0/490\\
97-121	&	cpl($\surd$)	&$1.12^{+0.31}_{-0.33}$	&$2.38^{+1.36}_{-0.68}$	&305.3/435& pl & 1.94$\pm$0.09 &  329.1/436\\
150-260	&	cpl($\surd$)	&$1.41^{+0.20}_{-0.21}$&$2.16^{+0.72}_{-0.46}$&403.9/577&  pl & $2.22^{+0.10}_{-0.06}$ &  460.4/578\\
260-505	&	cpl	&$2.14^{+0.14}_{-0.15}$	&$5.95^{+5.23}_{-1.96}$ &404.5/648& pl($\surd$) & 2.43$\pm$0.05 &  417.5/649\\
1172-1216&      cpl	&$1.19^{+0.31}_{-0.33}$	&$6.68^{+47.11}_{-3.26}$&290.4/508& pl($\surd$) & 1.53$\pm$0.10 & 293.9/509\\
5404-6082&      cpl($\surd$)	&1.21$\pm$0.05		&$7.33^{+1.33}_{-0.99}$	&783.3/821& pl & 1.52$\pm$0.02 & 904.2/822\\
11112-11883&    cpl($\surd$)	&1.68$\pm$0.10 		&$7.68^{+4.07}_{-2.02}$	&626.5/630& pl & 1.95$\pm$0.03 & 650.1/631\\
16347-17659&    cpl	&$1.90^{+0.31}_{-0.32}$	&$5.92^{+50.16}_{-2.91}$&248.5/489& pl($\surd$) & 2.22$\pm$0.10 & 251.8/490\\
22798-23598&	  .. & .. & ..&...	 &pl($\surd$)	&2.91$\pm$0.20 			&125.6/461	\\
28396-30083&      .. & .. & ..&...	 &pl($\surd$)	&$2.69^{+0.16}_{-0.15}$	&190.1/427\\
35290-76224&      .. & .. & ..&...	 &pl($\surd$)	&2.48$\pm$0.15&188.0/425\\
138272-201989&    .. & .. & ..&...	 &pl($\surd$)	&2.17$\pm$0.15&196.3/242\\
328262-485712&    .. & .. & ..&...	 &pl($\surd$)	&$2.52^{+0.21}_{-0.20}$&120.8/417\\
664241-2298128&   .. & .. & ..&...	 &pl($\surd$)	&$2.57^{+0.29}_{-0.27}$&113.5/319\\
\hline
\end{tabular}
\end{table}

\label{lastpage}
\end{document}